%% file: main.tex
\documentclass[10pt,conference]{IEEEtran}
\usepackage{cite}
\usepackage{amsmath,amssymb,amsfonts}
\usepackage{algorithmic}
\usepackage[ruled,vlined]{algorithm2e}
\usepackage{graphicx}
\usepackage{textcomp}
\usepackage{xcolor}
\usepackage{hyperref}
\usepackage{multirow}
\usepackage{multicol}
\usepackage{booktabs}
\usepackage{changes}
\definechangesauthor[name={joey}, color=red]{jl}

\definechangesauthor[name={kien}, color=blue]{kn}

\begin{document}

\title{Scaling Quantum Optimization for Unit Commitment via Pauli Correlation Encoding
}


\author{
\IEEEauthorblockN{
Kien X. Nguyen\IEEEauthorrefmark{1}\IEEEauthorrefmark{3},
Ilya Safro\IEEEauthorrefmark{1}\IEEEauthorrefmark{2},
Xiaoyuan Liu\IEEEauthorrefmark{3}
}
\IEEEauthorblockA{\IEEEauthorrefmark{1}\textit{Department of Computer and Information Sciences, University of Delaware}, Newark, DE, USA}
\IEEEauthorblockA{\IEEEauthorrefmark{2}\textit{Department of Physics and Astronomy, University of Delaware}, Newark, DE, USA}
\IEEEauthorblockA{\IEEEauthorrefmark{3}\textit{Quantum Computing Lab, Fujitsu Research of America}, Santa Clara, CA, USA}
\IEEEauthorblockA{\{kxnguyen, isafro\}@udel.edu, xliu@fujitsu.com}
}

\maketitle

\begin{abstract}
Unit commitment is an important optimization problem in power system operations, classified as NP-hard.
This paper presents a hybrid quantum–classical method for the unit commitment problem with time-dependent constraints, where decisions must be made about which generators to turn on/off and how much power they should produce over a planning horizon.
We use a hybrid quantum-classical optimization procedure to determine the on/off schedules of the generating units and the corresponding power dispatch that satisfies operational constraints such as load balance, generator limits, ramping, and reserve requirements.
We frame the optimization loop as a leader–follower structure, where the quantum optimizer leads to give the on/off decisions, and the classical optimizer follows to produce the power level schedule.
Leveraging Pauli-Correlation Encoding, our method scales to horizon-wide unit commitment schedules by encoding the binary variables with far fewer qubits.
By combining these components, the method can handle multi-period settings while using far fewer qubits than straightforward quantum encodings that allocate one qubit per decision variable as in prior approaches. 
We evaluate the approach on both small- and large-scale instances, up to 312 binary variables, and show that it reliably produces feasible schedules with competitive operating costs.
\end{abstract}


\input{sec/introduction}
\input{sec/formulation}
\input{sec/background}
\input{sec/method}
\input{sec/experiment}
\input{sec/conclusion}

\section*{Acknowledgment}
This work was supported in part by NSF award \#2444042.

\input{sec/appendix}


\clearpage
\bibliographystyle{IEEEtran}
\bibliography{main,ilya-biblio}

\end{document}

%% file: sec/introduction.tex
\section{Introduction}

Unit commitment (UC) is a fundamental optimization problem in power system operations, which aims to determine which generating units should be on or off and at what power levels they should operate over a planning horizon while satisfying demand and operational constraints~\cite{dokucomparative}.
UC is typically formulated as a mixed-integer optimization problem that combines binary commitment decisions with continuous power levels.
The complexity of the best known algorithm grows exponentially with the system size~\cite{guan2003optimization}, which renders the problem NP-hard~\cite{bendotti2019complexity}, and large-scale instances may involve thousands of decision variables and constraints. 

Although modern mixed-integer programming solvers have achieved impressive performance~\cite{knueven2020mixed,carrion2006computationally,chang2004practical,li2005price,ostrowski2011tight,putz2021comparison}, computational challenges remain for large systems, stochastic formulations, and real-time operations.
Recent advances in quantum computing have sparked interest in applying quantum algorithms to combinatorial optimization problems in such areas as finance \cite{herman2023quantum}, network science \cite{shaydulin2019network}, and drug discovery \cite{kumar2024recent} (to menion just a few). 
Variational quantum algorithms, such as the Quantum Approximate Optimization Algorithm (QAOA), have been proposed as promising candidates to tackle NP-hard optimization problems including some versions of the unit commitment  \cite{salgado2024hybrid,aboumrad2025}.

Despite growing interest in quantum optimization, applying quantum algorithms to \textit{realistic} UC problems remains largely unexplored.
Existing studies usually focus on simplified formulations or small benchmark instances due to hardware limitations and encoding challenges. 
For example, Koretsky et al.~\cite{koretsky2021} apply QAOA to a unit commitment formulation with a single time period. 
Aboumrad et al.~\cite{aboumrad2025} extend the setting to multiple periods, but still solve each time period independently, thereby ignoring inter-temporal constraints such as generator ramping limits~\cite{fan2002new}.

Recently, Salgado et al.~\cite{salgado2024hybrid} applied QAOA to a UC formulation that incorporates minimum up and down time constraints and spinning reserve requirements, resulting in a more realistic setting. However, their experiments remain limited (up to 12 generating units and 3 time periods) due to the computational restrictions of quantum circuit simulations.
A key challenge in applying quantum optimization to the unit commitment problem is scaling the formulation to realistic power system sizes, where practical instances often involve tens to hundreds of generating units across multiple time periods, resulting in a large number of binary commitment variables.
In conventional quantum formulations, each binary variable is mapped to a logical qubit, causing the required number of qubits, and consequently the cost of quantum simulation, to grow rapidly with system size.

\textbf{Our contribution:}  
We introduce a hybrid quantum–classical approach for scalable, time-dependent unit commitment that overcomes one of the main bottlenecks in quantum optimization for power systems: the qubit cost of encoding large multi-period commitment decisions. Our central idea is to use Pauli Correlation Encoding (PCE) ~\cite{Sciorilli_2025} to represent the full commitment schedule with far fewer qubits, making it possible to optimize the problem across the entire planning horizon at once. Unlike prior quantum approaches that simplify or decouple time periods, our framework directly incorporates inter-temporal constraints, including ramping, within the optimization loop.

We further couple this compressed quantum representation with a leader–follower hybrid architecture, where the quantum model (that acts as a leader) generates on/off schedules over full horizon and a classical dispatch optimizer (acts as a follower) computes the corresponding dispatch by choosing power levels that best satisfy load balance and operational constraints. 
This separation mirrors how UC is typically structured in practice, where discrete commitment decisions define the feasible operating region, and dispatch is optimized within that region. 
The resulting method scales beyond the small settings typically studied in prior work, and delivers feasible schedules with competitive costs on benchmark instances. In this sense, our work is a concrete step toward practical quantum optimization for realistic energy applications.


%% file: sec/formulation.tex
\section{Problem Formulation}

\subsection{Single-period Unit Commitment}

The Unit Commitment (UC) problem contains $N$ generator units. 
Each unit may be turned on or off, given by a binary variable $y_i\in\{0,1\}$. 
When $y_i=1$, the power level of the unit $p_i$ is a continuous real value constrained to $p^\text{min}_i\leq p_i \leq p^\text{max}_i$. 
When $y_i=0$, $p_i=0$. 
Together, these restrictions give an inequality constraint:
\[
p^\text{min}_iy_i \leq p_i \leq p^\text{max}_iy_i.
\]
The solution aims at (1) deciding which units to be turned on/off and (2) setting their corresponding power levels, so that the system satisfies a predefined load or demand, denoted as $L$.

Each power unit has a corresponding production cost function $F(y_i,p_i)$, which specifies the cost of turning on unit $i$ and for generating a certain amount of power $p_i$. This cost function is quadratic for UC:
\begin{align}
    F(y_i,p_i) = A_iy_i + B_ip_i + C_ip_i^2
    \label{eq:uc-obj}
\end{align}
where $A_i,B_i,C_i\in\mathbb{R}$ are constant. 
$A$ is the fixed cost coefficient that a unit necessarily incurs when it is turned on, regardless of the power it contributes. 
$B$ and $C$ are the linear and quadratic coefficients, respectively, and contribute to the unit's cost based on its power level. 
The objective for single-period unit commitment is defined as follows:
\begin{align}
    \min_{\mathbf{y},\mathbf{p}} &\sum_{i=1}^N F_i(y_i,p_i) \\
    \text{s.t. }&\sum_{i=1}^N p_i = L, \\
    &p^\text{min}_i\leq p_iy_i\leq p^\text{max}_i,\forall i\\
    &p_i\in\mathbb{R},y_i\in\{0,1\}, \forall i.
\end{align}

\subsection{Time-dependent Unit Commitment}

In order to more realistically model the UC problem, we must forbid units from turning on and shutting down abruptly for consecutive time periods~\cite{salgado2024hybrid}. We introduce a ramping mechanism using two predefined values $R_i^\text{up}$ and $R_i^\text{dn}$, which are the up- and down-ramping limits of unit $i$. Hence we have the $T$-period UC model as follows:
\begin{align}
    \min_{\mathbf{y},\mathbf{p}} &\sum_{t=1}^T\sum_{i=1}^N F_i(y_i^t,p_i^t) \\
    \text{s.t. }&\sum_{i=1}^N p_i^t = L^t,\quad \forall t,\\
    &p^\text{min}_iy_i^t\leq p_i^t\leq p^\text{max}_iy_i^t, \quad \forall i, t,\label{eq:cap-constraint}\\
    &\underbrace{p_i^{t+1} - p_i^t}_\text{up change} \leq R_i^\text{up}y_i^t, \quad\forall i, \forall t<T,\label{eq:simple-up-constraint}\\
    &\underbrace{p_i^t - p_i^{t+1}}_\text{down change} \leq R_i^\text{dn}y_i^{t+1}, \quad \forall i, \forall t<T,\label{eq:simple-down-constraint}\\
    & \sum_{i=1}^N p^\text{max}_iy_i^t \geq L^t + S^t, \forall t\\
    &p_i^t\in\mathbb{R},\quad y_i^t\in\{0,1\}, \forall i,t,
\end{align}
where Eq.~(\ref{eq:simple-up-constraint}) and~(\ref{eq:simple-down-constraint}) are the up and down ramping constraints, $S^t$ denotes the spinning reserve for time period $t$.
Note that the ramping constraints shown here are in simplified form and will be expanded and explained in more detail in Sec.~\ref{sec:method}.



%% file: sec/background.tex
\section{Background}
\subsection{Pauli Correlation Encoding}

Pauli Correlation Encoding (PCE) aims to compress $n_v$ variables into $n_q$ qubits, where $n_v \gg n_q$~\cite{Sciorilli_2025}.
PCE maps the variables in optimization problems to multi-body Pauli-matrix correlations, resulting in a polynomial compression of the problem's space requirements. 
By employing PCE, the number of qubits needed for encoding is reduced, making it particularly advantageous for near-term quantum devices with limited qubit resources. 
Furthermore, it is analytically demonstrated that PCE inherently mitigates barren plateaus, offering super-polynomial resilience against this phenomenon. 
This built-in feature enables unprecedented performance in quantum optimization solvers and has been applied to several constrained combinatorial optimization problems~\cite{soloviev2025large,padin2026pauli,do2026warm}.

\subsection{Mixed-Integer Quadratic Bilevel Optimization}
Mixed-integer quadratic bilevel optimization (MIQBLO) models hierarchical decisions where an upper-level mixed-integer choice parameterizes a lower-level convex quadratic program (QP)\cite{colson2007overview,dempe2002foundations,bard1998practical}. 
Unit commitment fits this template, where binary commitments define the operating region, and power dispatch is optimized by a quadratic program, giving us the power level generation schedule that has the least cost. 
Rather than solving MIQBLO exactly, we use a \textit{heuristic-guided bilevel approach} in which a variational quantum circuit proposes commitments and a classical QP solver computes the corresponding dispatch; circuit parameters are trained using bilevel sensitivities from the QP and gradients of quantum expectations \cite{Sciorilli_2025,osqp,Schuld_2019}.

%% file: sec/method.tex
\section{Method}

\label{sec:method}

\begin{figure*}
    \centering
    \includegraphics[width=\linewidth]{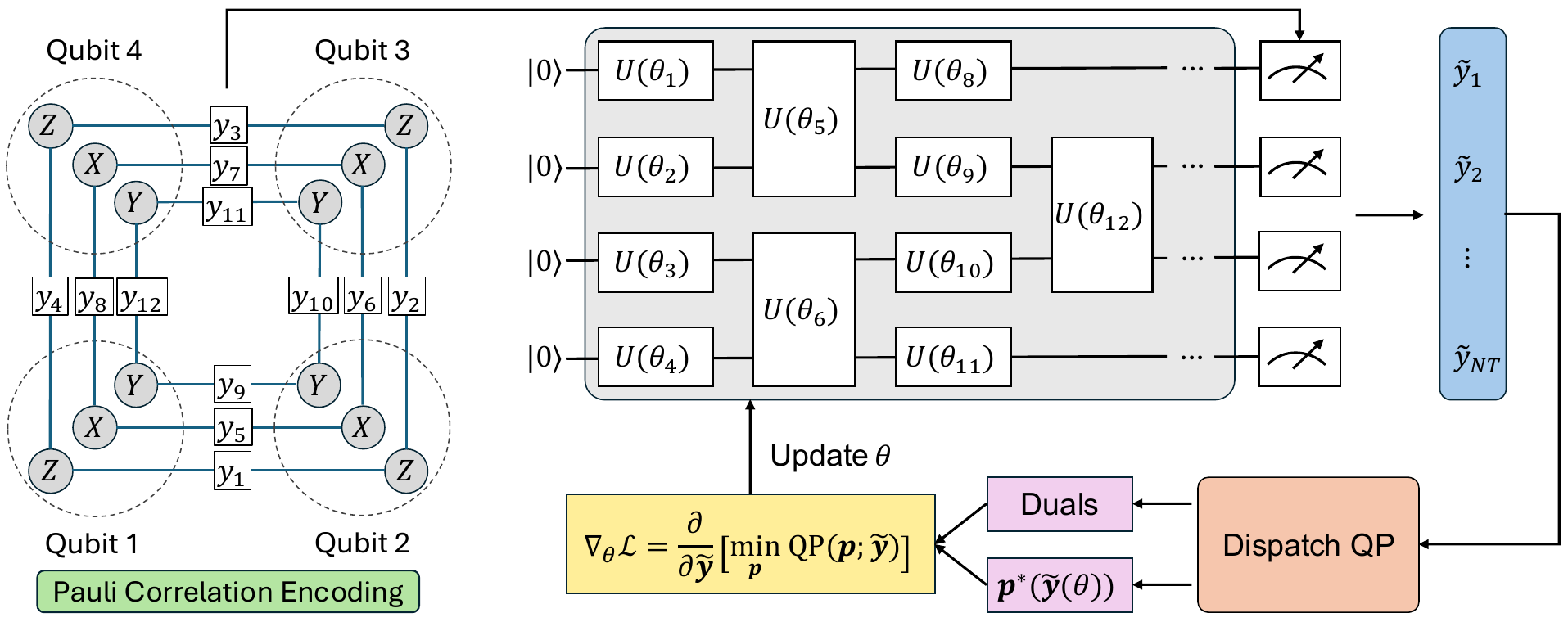}
    \caption{Overview of our heuristic-assisted differentiable bilevel optimization on a 12-variable problem, which is encoded into 4 qubits. The quantum circuit produces PCE correlators that decode to a soft commitment schedule $\mathbf{\tilde y}$. A classical QP computes the dispatch $\mathbf{p^*}$
    (and dual variables), which defines the training loss. Gradients are obtained by combining QP dual sensitivities with parameter-shift derivatives, and circuit parameters are updated with the Adam optimizer. After training, a threshold sweep produces a hard schedule and a final dispatch.}
    \label{fig:overview}
\end{figure*}

\subsection{Overview}
UC naturally admits a leader--follower structure: the leader chooses the binary commitment schedule $\mathbf y$, and the follower chooses a feasible dispatch $\mathbf p$ by solving a convex economic dispatch problem given $\mathbf y$.
Instead of solving the mixed-integer problem directly, we adopt a heuristic-assisted bilevel approach:
(i) a variational quantum circuit proposes a (soft) schedule $\tilde{\mathbf y}(\theta)\in[0,1]^{N\times T}$ using PCE, and
(ii) a classical convex solver computes a dispatch $\mathbf p^\star(\tilde{\mathbf y})$.
We then update circuit parameters $\theta$ using a bilevel gradient that combines sensitivities of the classical subproblem~\cite{bonnans2013perturbation,amos2017optnet,agrawal2019differentiable} with parameter-shift derivatives of quantum expectations~\cite{Mitarai_2018,Schuld_2019}.

First, we introduce Pauli Correlation Encoding (PCE) and describe the modifications required to represent time-indexed commitment decisions in our setting.
We then derive the lower-level objective used to compute the dispatch solution (power levels) given a commitment schedule. 
Finally, we derive the upper-level objective that couples the PCE-generated commitments with the dispatch subproblem and is used to optimize the variational quantum circuit parameters.

\subsection{Pauli-Correlation Encoding for Commitment Variables}
\label{subsec:pce_decode}

The conventional QAOA-style encoding would require one qubit per binary commitment variable $y_{i}^t$, leading to $NT$ qubits for a system with $N$ generating units over a horizon of $T$ time periods.
This linear scaling in both spatial and temporal dimensions quickly becomes impractical, as even moderately sized unit commitment instances lead to qubit requirements that exceed the capabilities of near-term quantum hardware.
We therefore adopt Pauli Correlation Encoding (PCE), which represents a larger number of logical binary variables using expectation values of multi-qubit Pauli correlations.

\paragraph{PCE observables and capacity}
Let $n_q$ be the number of qubits and let $k$ be the correlation order. PCE constructs a library of $k$-body Pauli strings by placing the same Pauli operator on a $k$-subset of qubits and identity elsewhere. In particular, we form three commuting families
\begin{align}
&\Pi^{(X)}=\{X_{a_1}\cdots X_{a_k}\},\nonumber \\
&\Pi^{(Y)}=\{Y_{a_1}\cdots Y_{a_k}\},\nonumber\\
&\Pi^{(Z)}=\{Z_{a_1}\cdots Z_{a_k}\}, 1\le a_1<\cdots<a_k\le n_q,
\end{align}
so that the total number of available correlators is
\begin{equation}
M_\text{max} \;=\; 3\binom{n_q}{k}.
\label{eq:pce-capacity}
\end{equation}
We choose the minimal $n_q$ such that $M_\text{max}\ge NT$ and select a subset of Pauli strings
\[
\{O_j\}_{j=1}^{NT}\subseteq \Pi^{(X)}\cup\Pi^{(Y)}\cup\Pi^{(Z)}
\] 
to represent all time dependent unit commitments. Each observable is assigned to a unique unit-time index via a fixed bijection $\varphi(j)=(i,t)$.

\paragraph{From correlators to soft commitments}
The three families $\Pi^{(X)},\Pi^{(Y)},\Pi^{(Z)}$ are each simultaneously measurable using a single measurement basis (global $X$, $Y$, or $Z$ basis, respectively). Thus, all $X$-type correlators are estimated from measurements in the $X$ basis (and similarly for $Y$ and $Z$), requiring only three global measurement settings per evaluation of the commitment schedule.
Given an ansatz $U(\theta)$ on $n_q$ qubits, we estimate the correlator expectations
\begin{equation}
e_j(\theta)\;=\;\langle 0|U(\theta)^\dagger O_j U(\theta)|0\rangle,\qquad \text{for }j=1,\dots,NT.
\end{equation}
In the original PCE definition, a hard logical bit is obtained by the sign rule $\mathrm{sgn}(e_j)\in\{-1,+1\}$. To enable continuous optimization and bilevel differentiation, we instead use a smooth surrogate based on a squashing nonlinearity:
\begin{equation}
\tilde y_{\varphi(j)}(\theta)\;=\;\frac{1+\tanh(\alpha\, e_j(\theta))}{2}\in[0,1],
\label{eq:pce-decode}
\end{equation}
where $\alpha>0$ controls the sharpness of the relaxation. Large $\alpha$ yields near-binary outputs, while smaller $\alpha$ provides a smoother training signal.

\paragraph{Hardening via thresholding}
After optimization, we convert the soft schedule to a binary commitment schedule by elementwise thresholding:
\begin{equation}
\hat y_{i,t}\;=\;\mathbb{I}[\tilde y_{i,t}(\theta)\ge\tau]\in\{0,1\},
\end{equation}
where $\mathbb{I}$ is the indicator function, and $\tau\in(0,1)$ is selected by a small grid search (Sec.~\ref{subsec:postprocess}) to obtain the best feasible schedule under the lower-level dispatch model.

\subsection{Lower-Level Objective Derivation}

\paragraph{Transition-aware ramping constraints}
For a unit that remains online across two consecutive periods, a ramp-up limit is often expressed as Eq.~\eqref{eq:simple-up-constraint} and similarly for ramp-down as Eq.~\eqref{eq:simple-down-constraint}.
However, Eq.~\eqref{eq:simple-up-constraint} alone does not specify how ramping interacts with startup and shutdown events.
To capture this interaction using linear constraints, we adopt the transition-aware ramp limits:
\begin{align}
p_i^{t+1}-p_i^t &\leq R_i^\text{up} y_i^t + p_i^\text{min}(y_i^{t+1}-y_i^t) + p_i^\text{max}(1-y_i^{t+1}),\nonumber\\
&\hspace{130pt}\forall i,\ \forall t<T,
\label{eq:full-up-constraint}\\
p_i^t - p_i^{t+1} &\leq R_i^\text{dn}y_i^{t+1} + p^\text{min}_i(y_i^t - y_i^{t+1}) + p^\text{max}_i(1-y_i^t),\nonumber\\
&\hspace{130pt}\forall i,\ \forall t<T.
\label{eq:full-down-constraint}
\end{align}
When a unit turns on ($y_i^t = 0,y_i^{t+1} = 1$), the ``up change" inequality~\eqref{eq:full-up-constraint} allows jumping to at least $p^\text{min}_{i}$ in one period (even if $R_i^\text{up} < p^\text{min}_{i}$).
When a unit stays on ($y_i^t = y_i^{t+1}=1$), the RHS of constraints \eqref{eq:full-up-constraint} and \eqref{eq:full-down-constraint} reduces to $R_i^\text{up}$ and $R_i^\text{dn}$.
When a unit shuts down ($y_i^t = 1,y_i^{t+1} = 0$), the ``down change" inequality~\eqref{eq:full-down-constraint} forces ramping to $p^\text{min}_{i}$ before shutting down.
When a unit stays off ($y_i^t = y_i^{t+1}=0$), the extra term deactivates the constraint.

\paragraph{Slack variables for lower-level optimization}
For arbitrary soft schedules $\tilde{\mathbf y}$, the dispatch constraints may become infeasible (e.g., insufficient committed capacity to meet load, or overly restrictive ramping).
To ensure the lower-level problem remains feasible and differentiable throughout optimization, we introduce slack variables and penalize them quadratically.
Specifically, we add (i) balance slack $s^\text{bal}_t$ to relax load balance, and (ii) ramp slacks $s^\text{up}_{i,t}$ and $s^\text{dn}_{i,t}$ to relax Eq.~\eqref{eq:full-up-constraint} and \eqref{eq:full-down-constraint}:
\begin{align}
\sum_{i=1}^{N} p_{i}^t + s^\text{bal}_t &= L^t, && \forall t,
\label{eq:slack-balance}\\
(p_i^{t+1}-p_i^t) - s^\text{up}_{i,t} &\le \mathrm{RHS}^{\text{up}}_{i,t}(\tilde{\mathbf y}), && \forall i,\ t<T,
\label{eq:slack-up}\\
(p_i^t-p_i^{t+1}) - s^\text{dn}_{i,t} &\le \mathrm{RHS}^{\text{dn}}_{i,t}(\tilde{\mathbf y}), && \forall i,\ t<T,
\label{eq:slack-down}
\end{align}
where $\mathrm{RHS}^{\text{up}}_{i,t}(\tilde{\mathbf y})$ and $\mathrm{RHS}^{\text{dn}}_{i,t}(\tilde{\mathbf y})$ are the right hand side of Eq.~\eqref{eq:full-up-constraint} and \eqref{eq:full-down-constraint}. Because the inequalities in Eq.~\eqref{eq:slack-up} and \eqref{eq:slack-down} become easier to satisfy as the slack variable increases, the optimizer will not choose negative value for the slack variable at the optimum, even without explicitly imposing $s\ge 0$.

\paragraph{Capacity gating with soft commitments}
$p_{i}^t$ is the power level for unit $i$ at time $t$ or \emph{dispatched output}. 
Thus, commitment decisions gate the feasible power range, per Eq.~\eqref{eq:cap-constraint}.
\begin{equation}
p_i^{\min}\,y_i^t \le p_i^t \le p_i^{\max}\,y_i^t,\qquad \forall i,t.
\label{eq:cap-gate}
\end{equation}
During training, the quantum model outputs $\tilde y_{i,t}\in[0,1]$, yielding a continuous relaxation of Eq.~\eqref{eq:cap-constraint} that remains linear in $(\mathbf{p},\tilde{\mathbf{y}})$.

\paragraph{Lower-level dispatch objective}
Given the relaxed unit commitment decisions $\tilde{\mathbf y}\in[0,1]^{N\times T}$, we compute the optimal power dispatch by solving a convex quadratic program that consolidates the constraint formulations described above. Let $\mathbf p\in\mathbb{R}^{N\times T}$ denote the generation schedule. The lower-level problem is defined as
\begin{align}
\mathbf p^\star(\tilde{\mathbf y}) = \arg\min_{\mathbf p,\mathbf s^\text{bal},\mathbf s^\text{up},\mathbf s^\text{dn}}\;\;
& \sum_{t=1}^{T}\sum_{i=1}^{N}\Big(B_i p_{i}^t+C_i {p_{i}^t}^2\Big)\nonumber\\
&+ \rho_\text{bal}\sum_{t=1}^{T}(s^\text{bal}_t)^2\nonumber\\
&+ \rho_\text{ramp}\sum_{i=1}^{N}\sum_{t=2}^{T}\Big((s^\text{up}_{i,t})^2+(s^\text{dn}_{i,t})^2\Big)
\label{eq:lower-obj}\\
&\text{s.t. }~\eqref{eq:slack-balance}~\eqref{eq:slack-up}~\eqref{eq:slack-down}~\eqref{eq:cap-gate}\nonumber,
\end{align}
where the objective combines quadratic generation costs with penalty terms that regulate violations of the previously defined balance and ramping constraints. The coefficients $\rho_\text{bal}$ and $\rho_\text{ramp}$ control the trade-off between cost minimization and adherence to these constraints under the relaxed decisions $\tilde{\mathbf y}$.

The resulting problem is a structured convex QP, which we solve efficiently using \textsc{OSQP}~\cite{osqp}. To accelerate convergence across bilevel iterations, we employ warm-starting by initializing each solve with the solution from the previous iteration.


\begin{algorithm}[h]
\caption{Differentiable Bilevel PCE--UC with Slackened Dispatch}
\label{alg:diffbilevel_pce_uc}
\KwIn{UC data; PCE observables $\{O_j\}$; ansatz $U(\theta)$; hyperparameters $(\alpha,\rho_\text{bal},\rho_\text{ramp},\lambda_\text{res})$; steps $n_s$.}
\KwOut{$\hat{\mathbf y}$, $\hat{\mathbf p}$.}
Initialize $\theta$\;
\For{$i=1$ \KwTo $n_s$}{
  Estimate $\{e_j(\theta)\}$ and decode $\tilde{\mathbf y}(\theta)$ via Eq.~\eqref{eq:pce-decode}\;
  Solve Eq.~\eqref{eq:lower-obj} to obtain $\mathbf p^\star$ and duals\;
  Compute bilevel gradient $\nabla_\theta \mathcal{J}(\theta)$ using dual sensitivities + parameter-shift\;
  Update $\theta$ with Adam optimizer\;
}
\textbf{Post-processing}

$\mathcal{T}\gets\{\tau_1,\ldots,\tau_k\}$\;
$\hat{\mathbf y}\gets \bot;\;\hat{\mathbf p}\gets \bot;\;J^\star\gets +\infty$\;

\ForEach{$\tau\in\mathcal{T}$}{
    Binarize $\mathbf y^{(\tau)} \gets \mathbb{I}[\tilde{\mathbf y}(\theta)\ge \tau]$\;
    Solve Eq.~\eqref{eq:lower-obj} to obtain $\mathbf p^{(\tau)} \gets \mathbf p^\star(\mathbf y^{(\tau)})$\;
    \If{$\mathbf p^{(\tau)}$ is feasible \textbf{and} $F(\mathbf y^{(\tau)},\mathbf p^{(\tau)}) < J^\star$}{
        $(\hat{\mathbf y},\hat{\mathbf p}) \gets (\mathbf y^{(\tau)},\mathbf p^{(\tau)})$\;
        $J^\star \gets F(\hat{\mathbf y},\hat{\mathbf p})$\;
    }
}
\Return $\hat{\mathbf y},\hat{\mathbf p}$\;
\end{algorithm}

\subsection{Upper-Level Objective and Spinning Reserve Penalty}
The upper-level objective evaluates the UC cost in the lower-level solution and includes a soft spinning-reserve penalty.
We define the reserve requirement $S^t$ and the headroom expression
\[
h_t(\tilde{\mathbf y})=\sum_i p_i^\text{max}\tilde y_{i}^t-(L^t+S^t),\forall t.
\]
We use a smooth penalty $\phi(\cdot)$ (e.g., softplus) to penalize spinning reserve shortfall:
\begin{equation}
\Phi_\text{res}(\tilde{\mathbf y})=\sum_{t=1}^{T}\phi\big(-h_t(\tilde{\mathbf y})\big)^2.
\label{eq:reserve-pen}
\end{equation}
The overall upper-level objective is
\begin{equation}
\min_{\theta}\;\; 
\mathcal{J}(\theta)=F\!\big(\tilde{\mathbf y}(\theta),\mathbf p^\star(\tilde{\mathbf y}(\theta))\big)
+\lambda_\text{res}\,\Phi_\text{res}\!\big(\tilde{\mathbf y}(\theta)\big),
\label{eq:upper-obj}
\end{equation}
where $\lambda_\text{res}$ is a tunable penalty coefficient.

\subsection{Dual sensitivity and parameter-shift gradients}
\label{subsec:gradients}

\paragraph{Dual variables and sensitivity}
The QP solver returns, in addition to the primal solution $(\mathbf p^\star,\mathbf s^{\text{bal}\star},\mathbf s^{\text{up}\star},\mathbf s^{\text{dn}\star})$, a vector of \emph{dual variables} (Lagrange multipliers) associated with the constraints in Eq.~\eqref{eq:slack-balance}--\eqref{eq:cap-gate}.
Intuitively, these multipliers quantify the marginal change in the optimal lower-level objective with respect to perturbations of the corresponding constraints.
In particular, since our capacity and ramping constraints depend on the leader decision $\tilde{\mathbf y}$ through their bounds, the associated dual variables provide the sensitivity of the optimal dispatch value with respect to $\tilde{\mathbf y}$ via the envelope theorem~\cite{bonnans2013perturbation}.
We use these sensitivities to compute $\nabla_{\tilde{\mathbf y}}\,F(\tilde{\mathbf y},\mathbf p^\star(\tilde{\mathbf y}))$ without differentiating through the iterative steps of the QP solver, enabling efficient bilevel updates of the circuit parameters~\cite{bonnans2013perturbation,amos2017optnet,agrawal2019differentiable}.

\paragraph{Quantum gradient via parameter shift}
For each circuit parameter $\theta_m$ that enters a single-parameter rotation with a $\{\pm 1\}$ generator, we compute $\partial e_j(\theta)/\partial \theta_m$ using the parameter-shift rule~\cite{crooks2019gradients,schuld2019evaluating}:
\begin{equation}
\frac{\partial e_j(\theta)}{\partial \theta_m} = \frac{1}{2}\Big(e_j(\theta_m+\tfrac{\pi}{2})-e_j(\theta_m-\tfrac{\pi}{2})\Big),
\end{equation}
and apply the chain rule through Eq.~\eqref{eq:pce-decode} to obtain $\nabla_\theta \mathcal{J}(\theta)$.
We then update $\theta$ using a first-order optimizer (e.g., Adam).

\subsection{Post-processing for Final Solution}
\label{subsec:postprocess}
After optimizing the quantum circuit parameters, the PCE decoder produces a soft commitment schedule $\tilde{\mathbf y}(\theta)\in[0,1]^{N\times T}$. 
We then apply elementwise thresholding with threshold $\tau\in[0,1]$, to obtain a binary schedule: 
$\mathbf y^{(\tau)}=\mathbb{I}[\tilde{\mathbf y}(\theta)\ge \tau]\in\{0,1\}^{N\times T}$.
Because the choice of $\tau$ can substantially affect feasibility and cost, we perform a grid search over a set of $K$ candidate thresholds $\mathcal{T}=\{\tau_k\}_{k=1}^{K}$ to select the best binary schedule.
For each $\tau\in\mathcal{T}$, we solve the lower-level dispatch problem to obtain $\mathbf p^{(\tau)}=\mathbf p^\star(\mathbf y^{(\tau)})$ and evaluate the UC objective $F(\mathbf y^{(\tau)},\mathbf p^{(\tau)})$. 
Among thresholds that yield a feasible dispatch, we select the final solution as
\begin{align}
\big(\mathbf y^{(\tau^\star)}, \mathbf p^{(\tau^\star)}\big),\; \text{where } \tau^\star = \arg\min_{\tau\in\mathcal{T}} \; F\!\big(\mathbf y^{(\tau)}, \mathbf p^{(\tau)}\big).
\end{align}
If no threshold produces a feasible solution, the experiment run is deemed infeasible.

%% file: sec/experiment.tex
\section{Results}
\subsection{Unit Commitment Instances}
\paragraph{Small-scale systems}
We evaluate our algorithm on five UC instances adopted from the work by Salgado et al.~\cite{salgado2024hybrid}.
Specifically, we use all but instance UC\_4a since there is no feasible solution for our formulation.
In the Appendix, we show the parameters of instance UC\_4b in Table~\ref{tab:4unit}, UC\_10a and UC\_10b in Table~\ref{tab:10unit}, and UC\_12a and UC\_12b in Table~\ref{tab:12unit_a} and~\ref{tab:12unit_b}, respectively.
The problem instances result in 12, 30 and 36 commitment variables.

\paragraph{Large-scale systems}
We adopt the 26-unit systems from Aboumrad et al.~\cite{aboumrad2025} and extend their single-period setting to a multi-period horizon by considering the first 12 time periods, resulting in a total of 312 binary commitment variables. 
As the original formulation does not include ramping constraints or spinning reserve requirements, we augment the system by synthesizing these constraints using parameter settings derived from the small-scale instances. 

In particular, we construct two systems, UC\_26a (easy) and UC\_26b (hard).
For UC\_26a, we generate relatively relaxed constraints by randomly sampling the up-ramping limits around 30\% and the down-ramping limits around 35\% of the maximum generation capacity $p_i^\text{max}$ for each unit. 
Additionally, the spinning reserve requirement is set to approximately 3\% of the total system load at each time period. 
For UC\_26b, the sampling range for ramping is tighter, around 20\% for up-ramping and 25\% for down-ramping; the spinning reverse requirement is around 5\%.
For UC\_26b, we impose tighter operational constraints by reducing the ramping limits, sampling the up-ramping coefficients around 20\% and the down-ramping coefficients around 25\% of $p_i^\text{max}$. 
The spinning reserve requirement is also increased to approximately 5\% of the load, resulting in a more constrained and challenging setting.

\subsection{Implementation Details}
We adopt Qiskit implementation of PCE~\cite{qiskitqce} to construct the variational quantum circuit and use OSQP~\cite{osqp} to solve the lower-level convex dispatch subproblem. 
Given $N$ units and $T$ time periods, we encode the $NT$ commitment variables using $k$-body Pauli correlators. 
Following the standard PCE capacity bound, we choose the minimum number of qubits $n_q$ such that $3\binom{n_q}{k}\ge NT$, and we split the correlators into three measurement groups (global $X$, $Y$, and $Z$ settings), requiring three measurement bases per evaluation of the schedule.

\begin{table*}[ht]
\centering
\caption{Results on small-scale systems with Brickwork and EfficientSU2 ansatzes.}
\resizebox{0.8\linewidth}{!}{%
\begin{tabular}{l|cc|ccc|c|c}
\toprule
\multirow{2}{*}{System} & \multirow{2}{*}{$n_v$} & \multirow{2}{*}{$n_q$} & \multirow{2}{*}{Best Cost} & \multirow{2}{*}{Mean Cost ($\pm$ Stdev)} & Feasibility & Optimal & Best \\
& & & & & Rate (\%)  & Cost &  Gap (\%)\\
\midrule
\multicolumn{8}{c}{\texttt{Brickwork}}\\
\midrule
UC\_4b & 12 & 4 & 32904.91 & 33787.49 ($\pm$ 674.80) & 80.0 & 32417.47 & 1.50\\
UC\_10a & 30 & 5 & 74055.35 & 74055.35 ($\pm$ 0.00) & 10.0 & 69070.09 & 7.21 \\
UC\_10b & 30 & 5 & 86400.47 & 88149.87 ($\pm$ 953.42) & 90.0 & 80447.49 & 7.39 \\
UC\_12a & 36 & 6 & 92507.71 & 93236.48 ($\pm$ 583.81) & 90.0 & 88070.83 & 5.03 \\
UC\_12b & 36 & 6 & 166697.83 & 168053.62 ($\pm$ 1356.75) & 60.0 & 154974.96 & 7.56\\
\midrule
\multicolumn{8}{c}{\texttt{EfficientSU2}}\\
\midrule
UC\_4b & 12 & 4 & 33641.97 & 34096.54 ($\pm$ 314.05) & 60.0 & 32417.47 & 3.77\\
UC\_10a & 30 & 5 & 72354.52 & 73098.41 ($\pm$ 579.18) & 30.0 & 69070.09 & 4.75 \\
UC\_10b & 30 & 5 & 85839.67 & 87215.42 ($\pm$ 1822.25) & 40.0 & 80447.49 & 6.70\\
UC\_12a & 36 & 6 & 91082.42 & 93429.23 ($\pm$ 1964.53) & 80.0 & 88070.83 & 3.41\\
UC\_12b & 36 & 6 & 163074.58 & 170217.14 ($\pm$ 5421.79) & 80.0 & 154974.96 & 5.22\\
\bottomrule
\end{tabular}%
}
\label{tab:results-main}
\end{table*}

\paragraph{Quantum ansatz}
We experiment with two parameterized circuit ansatze for PCE: (i) \texttt{brickwork} ansatz~\cite{cherrat2024quantum} consisting of repeated layers of single-qubit rotations followed by nearest-neighbor entanglers arranged in an even/odd pattern and (ii) Qiskit’s hardware-efficient \texttt{EfficientSU2}~\cite{efficientsu2} with linear entanglement.
Unless otherwise specified, we set the number of layers to 6, $k=2$ and $\alpha=(n_q)^2$.

\paragraph{Lower-level optimization}
We formulate power dispatch problem as a convex QP with a fixed sparse constraint matrix and iteration-dependent bounds that depend on the soft commitments $\tilde{\mathbf y}(\theta)$. To ensure feasibility during training, we introduce quadratic-penalty slack variables: $s^\text{bal}_t$ relaxes load balance and $s^\text{up}_{i,t},s^\text{dn}_{i,t}$ relax the ramp-up and ramp-down constraints. We warm-start OSQP with the previous iteration and cap its iterations per solve for efficiency. By default, $\rho_{\text{bal}}=10000$ and $\rho_{\text{ramp}}=1000$.

\paragraph{Upper-level optimization}
We optimize circuit parameters using a bilevel gradient that combines (i) sensitivity information returned by the QP solver (dual variables associated with constraints whose bounds depend on $\tilde{\mathbf y}$) and (ii) parameter-shift gradients for the quantum expectations.
To control the cost of parameter-shift evaluations, we update a subset of circuit parameters per iteration.
Unless otherwise specified, we run $n_s=200$ optimization steps and use a reserve-violation penalty with weight $\lambda_{\text{res}}=100$.

\subsection{Evaluation Metrics}
\label{subsec:metrics}

\paragraph{Feasibility rate}
Since the quantum circuit parameters are optimized via an unconstrained surrogate objective, we evaluate stability by repeating each experiment for 10 independent runs with different random seeds.
A run is counted as \emph{feasible} if the final post-processed schedule and dispatch $(\hat{\mathbf y},\hat{\mathbf p})$ satisfy all constraints. 
We report the feasibility rate as
\begin{equation}
\mathrm{FeasRate}(\%) = \frac{1}{10}\sum_{r=1}^{10}\mathbb{I}\big[(\hat{\mathbf y}^{(r)},\hat{\mathbf p}^{(r)})\ \text{is feasible}\big] \times 100,
\end{equation}
along with the average cost gap computed over feasible runs.

\paragraph{Operating cost}
To assess solution quality, we compare the operating cost obtained by our method against a reference solution computed by IBM ILOG CPLEX~\cite{cplex2022v22} on the same UC instance. 
Let $(\mathbf y^{\mathrm{cplex}},\mathbf p^{\mathrm{cplex}})$ denote the CPLEX solution and $F(\cdot,\cdot)$ the UC objective in \eqref{eq:uc-obj}. We calculate the normalized cost gap as
\begin{equation}
\mathrm{Gap} (\%) =
\frac{F(\hat{\mathbf y}^*,\hat{\mathbf p}^*) - F(\mathbf y^{\mathrm{cplex}},\mathbf p^{\mathrm{cplex}})}
{F(\mathbf y^{\mathrm{cplex}},\mathbf p^{\mathrm{cplex}})} \times 100,
\end{equation}
where $(\hat{\mathbf y}^*,\hat{\mathbf p}^*)$ is the result of the best run.


\paragraph{Violation percentage}
When deployed on large-scale systems, the proposed algorithm may produce solutions that violate some constraints due to the unconstrained nature of the upper-level optimization loop. Therefore, in addition to reporting objective values, we evaluate the degree of constraint violation to better characterize solution feasibility.

Let $\mathcal{C}$ denote the set of all constraints, and let $\mathcal{C}_\text{viol} \subseteq \mathcal{C}$ denote the subset of constraints that are violated under the post-processed commitment decision $\hat{\mathbf y}$ and dispatch solution $\mathbf p^\star$. We define the violation percentage as
\begin{align}
\mathrm{Violation}(\%) =
\frac{|\mathcal{C}_\text{viol}|}{|\mathcal{C}|} \times 100.
\end{align}

\begin{figure*}[ht]
    \centering
    \includegraphics[width=0.85\linewidth]{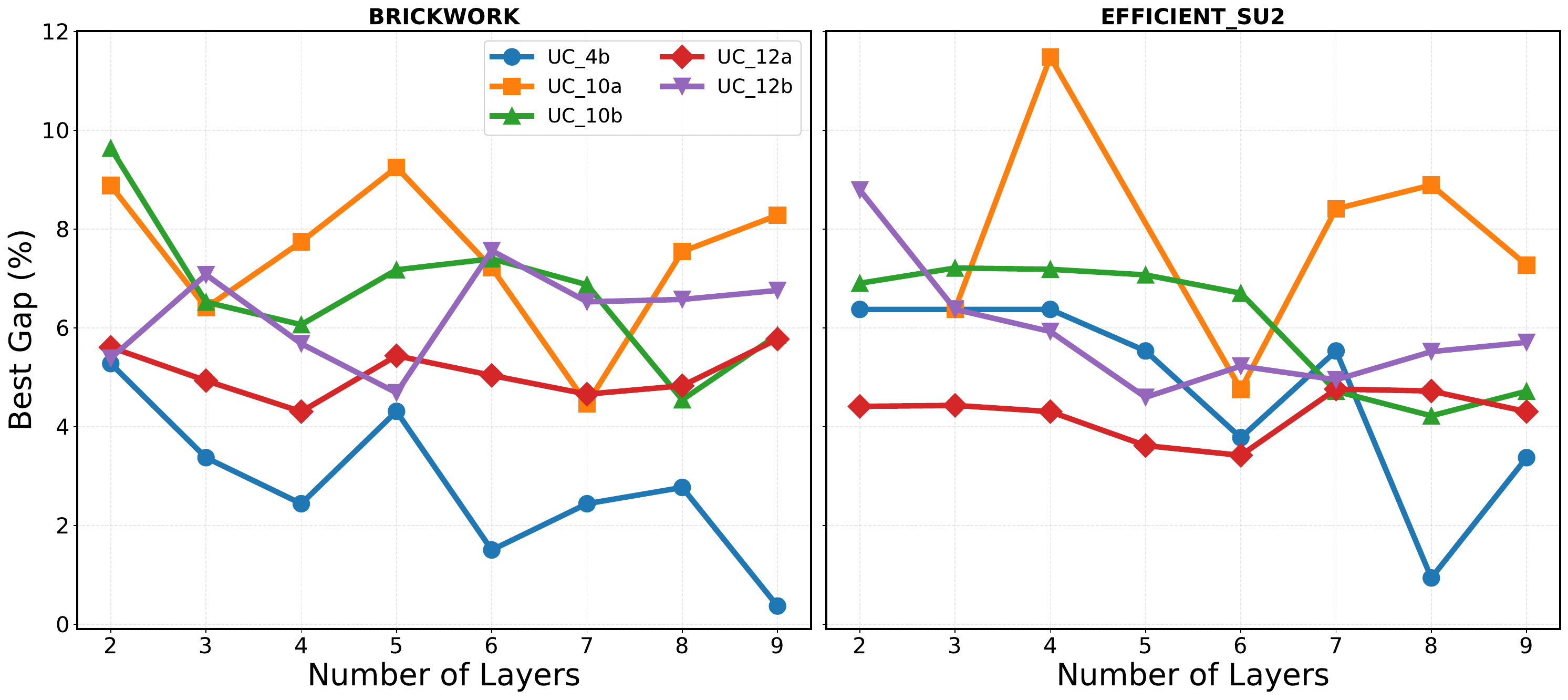}
    \caption{Ablation study on the number of ansatz layers over the range $\{2,3,\dots,9\}$ for both \texttt{brickwork} and \texttt{EfficientSU2} ansatzes. We report the best operating cost across the five small-scale systems.}
    \label{fig:ablate-layers-best}
\end{figure*}

\begin{figure*}[!h]
    \centering
    \includegraphics[width=0.85\linewidth]{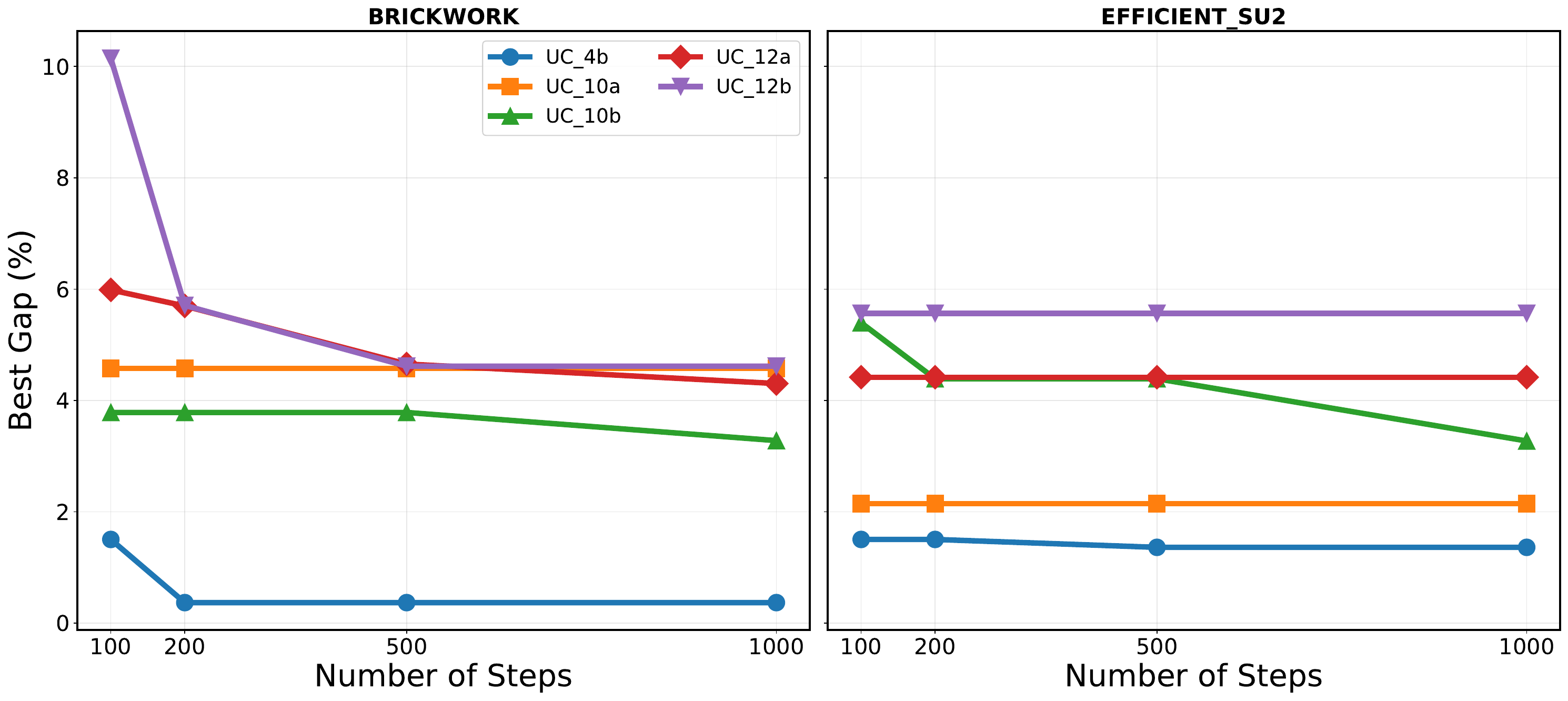}
    \caption{Ablation study on the number of optimization steps over the range $\{100,200,500,1000\}$ for both \texttt{brickwork} and \texttt{EfficientSU2} ansatzes. We report the best operating cost across the five small-scale systems.}
    \label{fig:ablate-steps-best}
\end{figure*}

\begin{figure}[!h]
    \centering
    \includegraphics[width=0.95\linewidth]{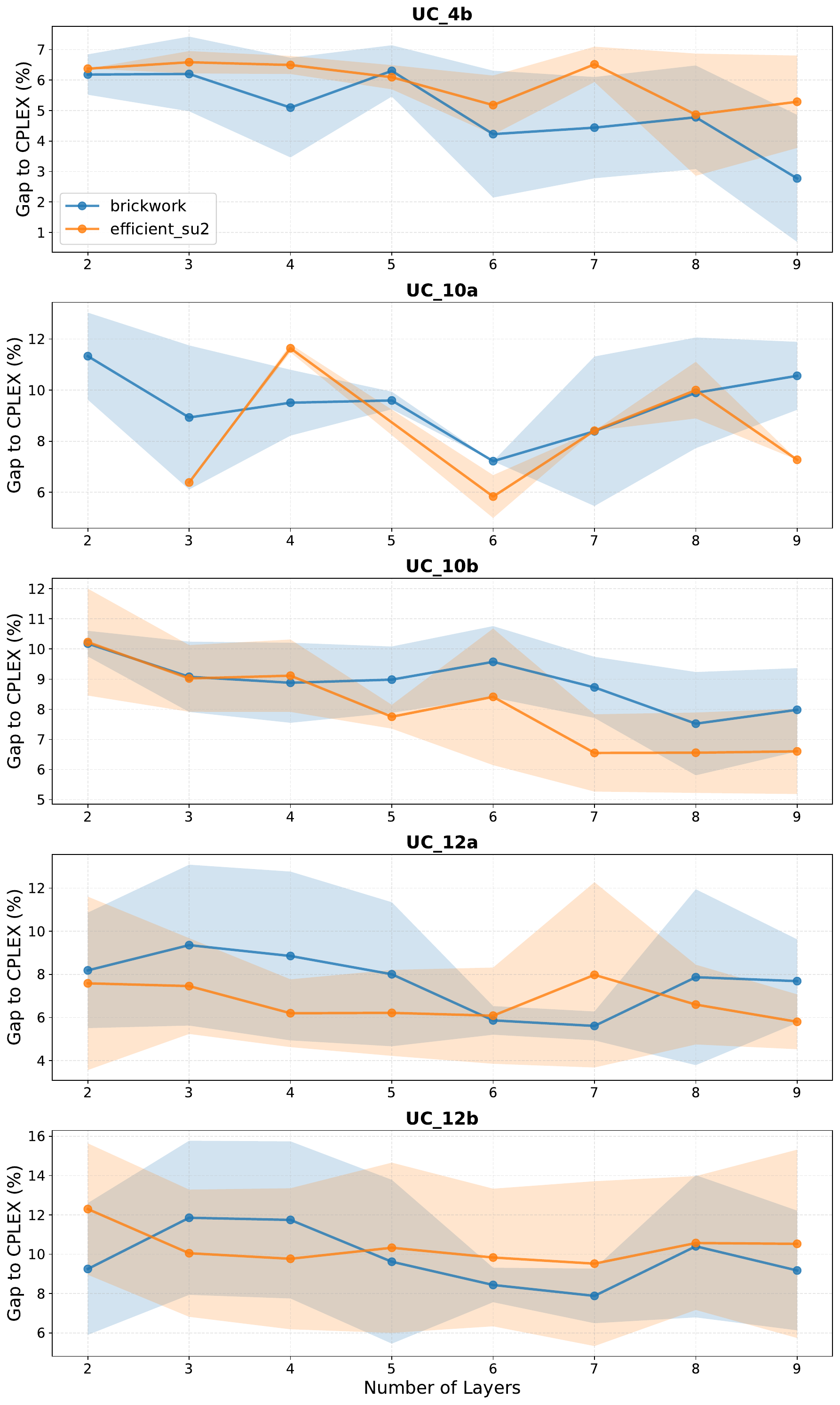}
    \caption{Ablation study on the number of ansatz layers over the range $\{2,3,\dots,9\}$ for both \texttt{brickwork} and \texttt{EfficientSU2} ansatzes. We report the average operating cost across the five small-scale systems.}
    \label{fig:ablate-layers-avg}
\end{figure}

This metric provides a normalized measure of how frequently the generated solution fails to satisfy the operational requirements. A lower violation percentage indicates that the method produces schedules that are not only cost-effective but also closer to feasible unit commitment solutions.

\begin{figure}[!h]
    \centering
    \includegraphics[width=0.95\linewidth]{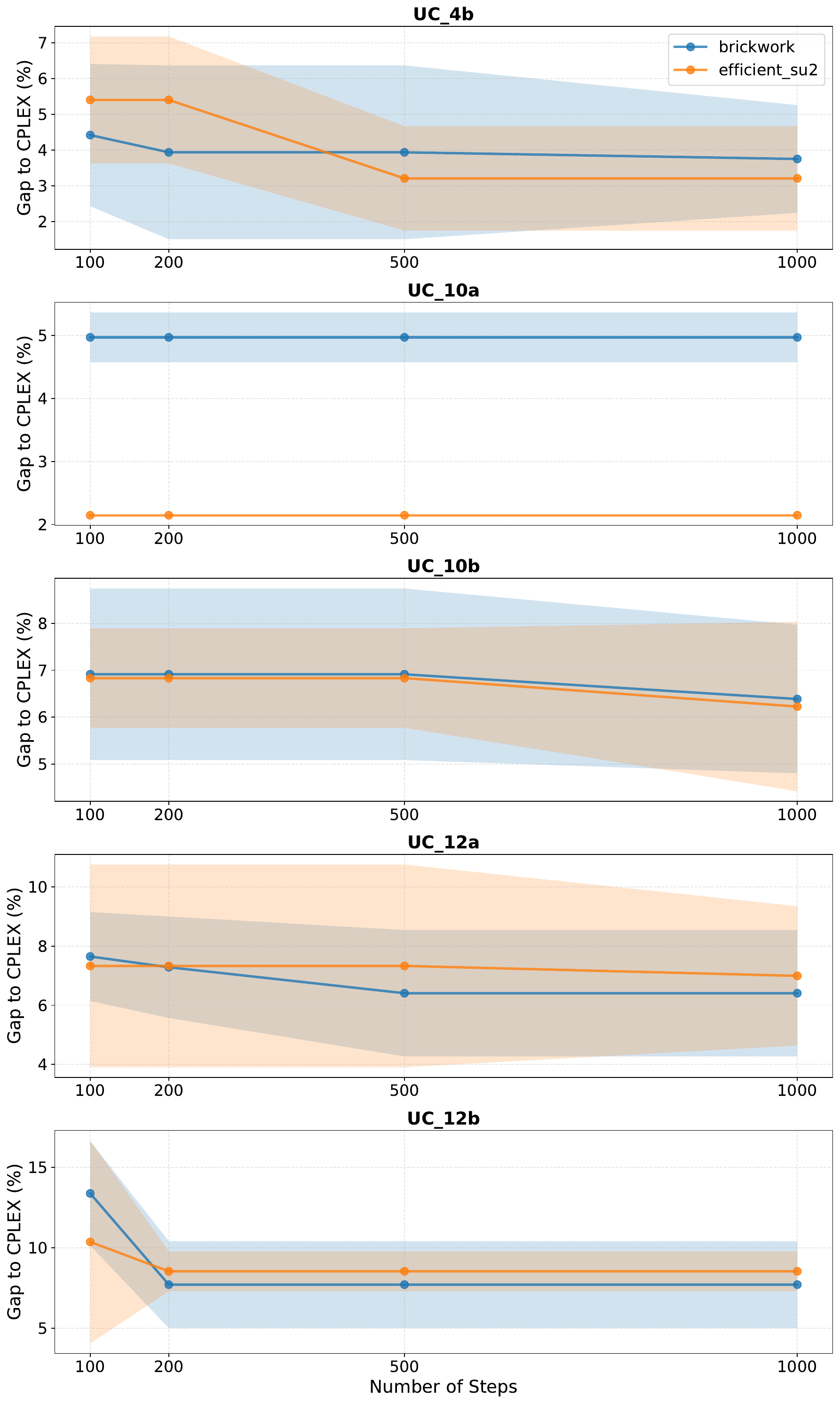}
    \caption{Ablation study on the number of optimization steps over the range $\{100,200,500,1000\}$ for both \texttt{brickwork} and \texttt{EfficientSU2} ansatzes. We report the average operating cost across the five small-scale systems.}
    \label{fig:ablate-steps-avg}
\end{figure}

\subsection{Results and Discussion on Small-scale Instances}

We report the results on small-scale instances for both ansatze, \texttt{brickwork} and \texttt{EfficientSU2}, in Table~\ref{tab:results-main}. 
Under the default hyper-parameter setting, \texttt{EfficientSU2} performs slightly better overall with an average best gap of 4.77\%, while \texttt{brickwork} attains 5.74\%.
However, \texttt{EfficientSU2} falls short in feasibility rate compared to \texttt{brickwork}, with an average deficit of 8\% over a total of 50 runs across the five instances.

A closer look reveals a trade-off between feasibility and cost quality. 
The \texttt{brickwork} ansatz achieves a higher feasibility rate on UC\_4b, UC\_10b, and UC\_12a compared to \texttt{EfficientSU2}, but especially struggles on UC\_10a where it only achieves a 10\% feasibility rate, or 1 successful run.
In contrast, \texttt{EfficientSU2}, despite having an overall lower feasibility rate, produces lower best objective values than those obtained by \texttt{brickwork} on the 10-unit and 12-unit systems. 
The only exception is UC\_4b, where \texttt{brickwork} achieves a close-to-optimal gap of 1.50\%.

The two ansatze also differ in stability across repeated runs. 
\texttt{brickwork} generally shows smaller standard deviations on the 10-unit and 12-unit systems, suggesting more consistent optimization behavior. 
For example, on UC\_10b, the standard deviation of \texttt{brickwork} is 953.42 compared to 1822.25 for \texttt{EfficientSU2}. 
A similar pattern appears on UC\_12b, where \texttt{brickwork} reduces the standard deviation from 5421.79 to 1356.75. 
On the other hand, \texttt{EfficientSU2} remains competitive in mean cost and attains the best overall average gap, indicating that its search space may be more expressive but also harder to optimize reliably.

Overall, these results suggest that \texttt{EfficientSU2} is preferable when the primary goal is obtaining the lowest possible cost, whereas \texttt{brickwork} is more attractive when robustness and feasibility are prioritized. 
This trade-off is especially relevant in unit commitment, where infeasible samples cannot be used in practice. 
The results also show that, even with only 4--6 qubits for instances containing 12--36 logical commitment variables, PCE is able to recover solutions within roughly 1--7\% of the CPLEX optimum on all benchmark systems, demonstrating the promise of the proposed framework for scaling quantum optimization to larger UC formulations.

\subsection{Ablation Studies on Small-scale Systems}

\paragraph{Ablation on Ansatz Depth}
In this experiment, we investigate the impact of ansatz depth on the performance of the proposed algorithm. Specifically, we vary the number of ansatz layers over $\{2,3,\dots,9\}$ for both \texttt{brickwork} and \texttt{EfficientSU2}, and repeat each configuration over 10 independent runs. 
Across the five small-scale systems, we report the best operating cost in Figure~\ref{fig:ablate-layers-best} and the mean operating cost with standard deviation in Figure~\ref{fig:ablate-layers-avg}.

According to Figure~\ref{fig:ablate-layers-best}, for both ansatz types, we observe a clear downward trend in operating cost as the number of layers increases on UC\_4b and UC\_10b, suggesting that additional circuit depth improves expressivity for these instances. 
Notably, on UC\_4b, \texttt{brickwork} achieves a gap of only 0.36\% from the optimal solution at depth 10. 
For the 12-unit systems, the operating cost initially decreases as the ansatz depth increases, reaching its best performance around layers 4--5 for \texttt{brickwork} and around layers 5--6 for \texttt{EfficientSU2}. 
Beyond this point, however, the cost begins to increase on UC\_12a and UC\_12b, indicating that deeper circuits do not necessarily translate into better solutions and may introduce additional optimization difficulty.
For the UC\_10a system, which is particularly challenging as evidenced by the low feasibility rates reported in Table~\ref{tab:results-main}, we observe significant fluctuations in performance across both ansatzes. 
In particular, \texttt{EfficientSU2} fails to produce any feasible solutions at depths 2 and 5, highlighting the sensitivity of this instance to the choice of ansatz and circuit depth.

\begin{table*}[h]
\centering
\caption{Results on large-scale systems UC\_26a and UC\_26b with brickwork ansatz.}
\resizebox{0.85\linewidth}{!}{%
\begin{tabular}{l|c|cc|ccc|ccc}
\toprule
\multirow{2}{*}{System} & \multirow{2}{*}{Run} & \multirow{2}{*}{$n_v$} & \multirow{2}{*}{$n_q$} & \multirow{2}{*}{Cost} & Optimal & Gap & Num. of & Num. of & Violation \\
& & & &  & Cost & (\%) & Constraints & Violations & Percentage (\%)\\
\midrule
\multirow{3}{*}{UC\_26a} & 1 & 312 & 15 & 358981.44 & 312510.44 & 14.87 & 1220 & 88 & 7.21 \\
& 2 & 312 & 15 & 355061.29 & 312510.44 & 13.61 & 1220 & 0 & 0.00\\
& 3 & 312 & 15 & 359422.96 & 312510.44 & 15.01 & 1220 & 92 & 7.54 \\
\midrule
\multirow{3}{*}{UC\_26b} & 1 & 312 & 15 & 363117.57 & 314692.81 & 15.38 & 1220 & 98 & 8.03 \\
& 2 & 312 & 15 & 364997.67 & 314692.81 & 15.98 & 1220 & 73 & 5.98 \\
& 3 & 312 & 15 & 356532.11 & 314692.81 & 13.29 & 1220 & 79 & 6.47 \\
\bottomrule
\end{tabular}%
}
\label{tab:results-largescale}
\end{table*}

\paragraph{Ablation on Optimization Steps}
In this experiment, we investigate the impact of the number of optimization steps on the performance of the proposed algorithm. 
Specifically, we vary the optimization budget as $n_s \in \{100, 200, 500, 1000\}$ while fixing the ansatz depth to 6. 
During optimization, we record the best solution found at each checkpoint in $\{100, 200, 500, 1000\}$ to evaluate how solution quality evolves as the optimization budget increases.
Similar to above, we report the best operating cost in Figure~\ref{fig:ablate-steps-best} and the average operating cost with standard deviation in Figure~\ref{fig:ablate-steps-avg}.

According to Figure~\ref{fig:ablate-steps-best}, the operating cost generally decreases as the number of optimization steps increases, indicating that larger optimization budgets improve solution quality. 
For the 12-unit systems, \texttt{brickwork} appears to converge around 500 steps on UC\_12b and around 1000 steps on UC\_12a, whereas \texttt{EfficientSU2} reaches its best observed performance as early as 100 steps. 
For UC\_10b, the operating cost continues to decrease up to 1000 steps for both ansatzes, suggesting that this instance may require a larger optimization budget to fully converge.
A similar trend is observed in Figure~\ref{fig:ablate-steps-avg}, where the mean operating cost decreases as the number of optimization steps increases. 
In addition, the standard deviation becomes smaller for the 4-unit and 12-unit systems, indicating that longer optimization runs not only improve the best-found solutions but also lead to more stable performance across runs.

\subsection{Results and Discussion on Large-scale Instances}
In this section, we evaluate the proposed algorithm on large-scale 26-unit systems over a 12-period planning horizon, resulting in 312 binary commitment variables. 
Through Pauli Correlation Encoding, this problem is represented using only 15 qubits, substantially reducing the qubit requirement compared to conventional one-qubit-per-variable encodings. 
We use the \texttt{brickwork} ansatz with 8 layers, as it exhibits more stable performance on the small-scale systems. 
For each large-scale instance, we optimize the quantum circuit for 500 steps and perform 3 independent runs.
For each run, we report the operating cost, optimality gap with respect to the CPLEX benchmark, and violation percentage in Table~\ref{tab:results-largescale}.

The proposed method is able to produce competitive solutions using only 15 qubits, demonstrating its scalability under the PCE framework. 
For UC\_26a, we observe variability across runs: while two runs incur constraint violations (around 7\%), one run achieves a fully feasible solution with zero violations and a gap of 13.61\%. 
This indicates that the method is capable of finding feasible solutions, although feasibility is not consistent due to the unconstrained optimization process. 

For the more challenging UC\_26b system, all runs exhibit nonzero violation percentages (ranging from 5.98\% to 8.03\%), reflecting the tighter operational constraints. 
The optimality gaps remain in a similar range, 13.29\%--15.98\%, suggesting that while the algorithm maintains stable cost performance, it struggles more with feasibility under stricter conditions. 
Overall, the method scales to large instances with significantly reduced qubit requirements, but may require additional mechanisms to consistently enforce constraint satisfaction.

\begin{table}[!h]
\centering
\caption{Warmstart speedup on large-scale systems UC\_26a and UC\_26b.}
\resizebox{0.85\linewidth}{!}{%
\begin{tabular}{l|ccc}
\toprule
System & Warmstart & No Warmstart & Speedup (\%) \\
\midrule
UC\_26a & 0.10 sec & 0.12 sec & 16.67 \\
UC\_26b & 0.63 sec & 0.66 sec & 4.54 \\
\bottomrule
\end{tabular}%
}
\label{tab:results-largescale}
\end{table}

\subsection{Warm-start for Classical Solver}

We evaluate the benefit of using quantum-generated solutions to warm-start the classical CPLEX solver on large-scale instances. 
As shown in Table~\ref{tab:results-largescale}, warm-starting consistently reduces the solve time compared to running CPLEX from scratch. 
For UC\_26a, we observe a speedup of 16.67\%, reducing the runtime from 0.12 seconds to 0.10 seconds. A smaller but still consistent improvement is observed on UC\_26b, with a speedup of 4.54\%.
The results suggest that the quantum-generated solutions provide useful initial incumbents that guide the branch-and-bound process more effectively, allowing CPLEX to prune the search space earlier. 

%% file: sec/conclusion.tex
\section{Conclusion}

In this work, we propose a hybrid quantum-classical method for the time-dependent unit commitment problem, considering more realistic and complex constraints compared to existing works. We use a leader–follower optimization structure, where the quantum optimizer makes the unit on/off decisions, and the classical optimizer follows to produce the power generation schedule.
In addition, by leveraging Pauli-Correlation Encoding, our method has a polynomial reduction on number of qubits needed compared to conventional methods, making it advantageous for near-term quantum devices with a limited number of qubits.
Combining these components, we show that our method can reliably produce feasible schedules with competitive operating costs compared to classical solver on standard instances. This opens a path towards solving industry-level large-scale optimization with complex constraints using quantum optimization, highlighting the possibilities of achieving quantum advantage in practical applications. 

\noindent\textbf{Future Work.} A natural direction for future work is to extend the proposed framework toward industry-grade deployment, where unit commitment instances typically involve substantially larger numbers of generating units, longer planning horizons, transmission-network constraints, contingency requirements, and uncertainty arising from renewable generation and load variability. In such settings, further scalability will likely require the integration of a very large-scale decomposition strategy, such as multilevel approach for QAOA \cite{bach2024mlqaoa,angone2023hybrid,maciejewski2024multilevel}, to divide the global commitment problem into coordinated subproblems while preserving temporal and operational consistency. Another plausible future direction is to train the quantum model so that its parameters transfer across related unit commitment instances, such as different load profiles or reserve levels. This could reduce the need to optimize from scratch for every new case and improve scalability in repeated or decomposition-based solves. This can be done in the spirit of trained models for QAOA \cite{falla2024graph,galda2023similarity}. Another promising direction is to develop a QAOA-GPT style framework \cite{tyagin2025qaoa} for unit commitment, in which a generative model predicts high-quality parameters from instance features such as load trajectories, reserve requirements, and generator characteristics. Such an approach could substantially reduce per-instance training cost and provide transferable warm starts for large-scale or decomposition-based unit commitment problems.

%% file: sec/appendix.tex
\appendices
\section{Small-scale Instances}

Table~\ref{tab:4unit}--\ref{tab:12unit_b} show the details about small-scale instances.

\begin{table}[h]
\centering
\caption{Parameters for the UC\_4b system.}
\resizebox{0.9\linewidth}{!}{%
\begin{tabular}{c|ccccccc}
\toprule
Unit & $A$ & $B$ & $C$ & $p_\text{min}$ & $p_\text{max}$ & $R^\text{up}$ & $R^\text{dn}$ \\
\midrule
1 & 1000 & 16.19 & 0.00048 & 150 & 455 & 80 & 100 \\
2 & 700  & 16.50 & 0.00200 & 20  & 130 & 15 & 30 \\
3 & 450  & 16.70 & 0.00398 & 25  & 165 & 30 & 40 \\
4 & 370  & 22.26 & 0.00712 & 20  & 80  & 5  & 10 \\
\toprule
\multicolumn{2}{l|}{$t$} & \multicolumn{2}{c|}{1} & \multicolumn{2}{c|}{2} & \multicolumn{2}{c}{3} \\
\midrule
\multicolumn{2}{l|}{$L$} & \multicolumn{2}{c|}{650} & \multicolumn{2}{c|}{530} & \multicolumn{2}{c}{450} \\
\multicolumn{2}{l|}{$S$} & \multicolumn{2}{c|}{50} & \multicolumn{2}{c|}{25} & \multicolumn{2}{c}{15} \\
\bottomrule
\end{tabular}%
}
\label{tab:4unit}
\end{table}

\begin{table}[h]
\centering
\caption{Parameters for the UC\_10a and UC\_10b systems.}
\resizebox{0.9\linewidth}{!}{%
\begin{tabular}{c|ccccccc}
\toprule
Unit & $A$ & $B$ & $C$ & $p_\text{min}$ & $p_\text{max}$ & $R^\text{up}$ & $R^\text{dn}$ \\
\midrule
1  & 660  & 25.92 & 0.00413 & 10  & 55  & 80 & 25 \\
2  & 670  & 27.76 & 0.00173 & 10  & 55  & 20 & 10 \\
3  & 700  & 16.60 & 0.00200 & 20  & 130 & 20 & 30 \\
4  & 680  & 16.50 & 0.00211 & 20  & 130 & 40 & 50 \\
5  & 450  & 19.70 & 0.00398 & 25  & 165 & 35 & 35 \\
6  & 970  & 17.26 & 0.00031 & 150 & 455 & 50 & 60 \\
7  & 480  & 27.74 & 0.00790 & 25  & 85  & 15 & 70 \\
8  & 665  & 27.27 & 0.00222 & 10  & 55  & 80 & 100 \\
9  & 1000 & 16.19 & 0.00048 & 150 & 455 & 50 & 80 \\
10 & 370  & 22.26 & 0.00712 & 20  & 80  & 30 & 40 \\
\toprule
\multicolumn{2}{l|}{$t$} & \multicolumn{2}{c|}{1} & \multicolumn{2}{c|}{2} & \multicolumn{2}{c}{3} \\
\midrule
\multicolumn{2}{l|}{$L$ (UC\_10a)} & \multicolumn{2}{c|}{900} & \multicolumn{2}{c|}{1000} & \multicolumn{2}{c}{1300} \\

\multicolumn{2}{l|}{$L$ (UC\_10b)} & \multicolumn{2}{c|}{1300} & \multicolumn{2}{c|}{1400} & \multicolumn{2}{c}{1200} \\
\midrule
\multicolumn{2}{l|}{$S$} & \multicolumn{2}{c|}{20} & \multicolumn{2}{c|}{10} & \multicolumn{2}{c}{30} \\
\bottomrule
\end{tabular}%
\label{tab:10unit}
}
\end{table}

\begin{table}[h]
\centering
\caption{Parameters for the UC\_12a system.}
\resizebox{0.9\linewidth}{!}{%
\begin{tabular}{c|ccccccc}
\toprule
Unit & $A$ & $B$ & $C$ & $p_\text{min}$ & $p_\text{max}$ & $R^\text{up}$ & $R^\text{dn}$ \\
\midrule
1  & 660  & 25.92 & 0.00413 & 10  & 55  & 80 & 25 \\
2  & 670  & 27.76 & 0.00173 & 10  & 55  & 20 & 10 \\
3  & 700  & 16.60 & 0.00200 & 20  & 130 & 20 & 30 \\
4  & 680  & 16.50 & 0.00211 & 20  & 130 & 40 & 50 \\
5  & 450  & 19.70 & 0.00398 & 25  & 165 & 35 & 35 \\
6  & 970  & 17.26 & 0.00031 & 150 & 455 & 50 & 60 \\
7  & 480  & 27.74 & 0.00790 & 25  & 85  & 15 & 70 \\
8  & 665  & 27.27 & 0.00222 & 10  & 55  & 80 & 100 \\
9  & 1000 & 16.19 & 0.00048 & 150 & 455 & 50 & 80 \\
10 & 370  & 22.26 & 0.00712 & 20  & 80  & 30 & 40 \\
11 & 490  & 18.50 & 0.00740 & 50  & 185 & 70 & 40 \\
12 & 735  & 24.90 & 0.00154 & 120 & 370 & 60 & 80 \\
\toprule
\multicolumn{2}{l|}{$t$} & \multicolumn{2}{c|}{1} & \multicolumn{2}{c|}{2} & \multicolumn{2}{c}{3} \\
\midrule
\multicolumn{2}{l|}{$L$} & \multicolumn{2}{c|}{1500} & \multicolumn{2}{c|}{1350} & \multicolumn{2}{c}{1450} \\
\multicolumn{2}{l|}{$S$} & \multicolumn{2}{c|}{20} & \multicolumn{2}{c|}{10} & \multicolumn{2}{c}{30} \\
\bottomrule
\end{tabular}%
}
\label{tab:12unit_a}
\end{table}

\begin{table}[h]
\centering
\caption{Parameters for the UC\_12b system.}
\resizebox{0.9\linewidth}{!}{%
\begin{tabular}{c|ccccccc}
\toprule
Unit & $A$ & $B$ & $C$ & $p_\text{min}$ & $p_\text{max}$ & $R^\text{up}$ & $R^\text{dn}$ \\
\midrule
1  & 960  & 20.40 & 0.00287 & 170 & 355 & 40  & 75 \\
2  & 470  & 29.80 & 0.00788 & 20  & 55  & 30  & 60 \\
3  & 560  & 28.50 & 0.00646 & 85  & 400 & 50  & 40 \\
4  & 400  & 15.90 & 0.00057 & 155 & 360 & 70  & 35 \\
5  & 600  & 27.90 & 0.00260 & 195 & 430 & 30  & 85 \\
6  & 1000 & 17.20 & 0.00584 & 200 & 465 & 40  & 70 \\
7  & 900  & 17.70 & 0.00199 & 100 & 275 & 70  & 75 \\
8  & 910  & 27.30 & 0.00454 & 65  & 305 & 80  & 50 \\
9  & 830  & 21.30 & 0.00270 & 15  & 70  & 60  & 85 \\
10 & 750  & 24.40 & 0.00150 & 160 & 320 & 100 & 30 \\
11 & 860  & 28.90 & 0.00260 & 30  & 220 & 50  & 80 \\
12 & 980  & 21.90 & 0.00109 & 60  & 470 & 70  & 65 \\
\toprule
\multicolumn{2}{l|}{$t$} & \multicolumn{2}{c|}{1} & \multicolumn{2}{c|}{2} & \multicolumn{2}{c}{3} \\
\midrule
\multicolumn{2}{l|}{$L$} & \multicolumn{2}{c|}{2000} & \multicolumn{2}{c|}{2200} & \multicolumn{2}{c}{2500} \\
\multicolumn{2}{l|}{$S$} & \multicolumn{2}{c|}{50} & \multicolumn{2}{c|}{20} & \multicolumn{2}{c}{40} \\
\bottomrule
\end{tabular}%
}
\label{tab:12unit_b}
\end{table}

\input{sec/large-systems}

%% file: sec/large-systems.tex
\clearpage
\section{Large-scale Instances}
Table~\ref{tab:26unit_a} and \ref{tab:26unit_b} show the details about the large-scale 26-unit instances.

\begin{table}[h]
\centering
\caption{Parameters for the 26unit\_time\_a system.}
\resizebox{\linewidth}{!}{%
\begin{tabular}{c|ccccccc}
\toprule
Unit & $A$ & $B$ & $C$ & $p_{\text{min}}$ & $p_{\text{max}}$ & $R^{\text{up}}$ & $R^{\text{dn}}$ \\
\midrule
1 & 24.3891 & 25.55 & 0.02533 & 2.40 & 12.00 & 4 & 5 \\
2 & 24.4110 & 25.68 & 0.02649 & 2.40 & 12.00 & 4 & 5 \\
3 & 24.6382 & 25.80 & 0.02801 & 2.40 & 12.00 & 4 & 5 \\
4 & 24.7605 & 25.93 & 0.02842 & 2.40 & 12.00 & 4 & 5 \\
5 & 24.8882 & 26.06 & 0.02855 & 2.40 & 12.00 & 4 & 5 \\
6 & 117.7550 & 37.55 & 0.01199 & 4.00 & 20.00 & 6 & 8 \\
7 & 118.1080 & 37.66 & 0.01261 & 4.00 & 20.00 & 6 & 8 \\
8 & 118.4580 & 37.78 & 0.01359 & 4.00 & 20.00 & 6 & 8 \\
9 & 118.8210 & 37.89 & 0.01433 & 4.00 & 20.00 & 6 & 8 \\
10 & 81.1364 & 13.33 & 0.00876 & 15.20 & 76.00 & 23 & 28 \\
11 & 81.2980 & 13.36 & 0.00895 & 15.20 & 76.00 & 23 & 28 \\
12 & 81.4641 & 13.38 & 0.00910 & 15.20 & 76.00 & 23 & 28 \\
13 & 81.6259 & 13.41 & 0.00932 & 15.20 & 76.00 & 23 & 28 \\
14 & 217.8950 & 18.00 & 0.00623 & 25.00 & 100.00 & 30 & 35 \\
15 & 218.3350 & 18.10 & 0.00612 & 25.00 & 100.00 & 30 & 35 \\
16 & 218.7750 & 18.20 & 0.00598 & 25.00 & 100.00 & 30 & 35 \\
17 & 142.7350 & 10.69 & 0.00463 & 54.25 & 155.00 & 47 & 54 \\
18 & 142.0290 & 10.72 & 0.00473 & 54.25 & 155.00 & 47 & 54 \\
19 & 143.3180 & 10.74 & 0.00481 & 54.25 & 155.00 & 47 & 54 \\
20 & 143.5970 & 10.76 & 0.00487 & 54.25 & 155.00 & 47 & 54 \\
21 & 259.1310 & 23.00 & 0.00259 & 68.95 & 197.00 & 59 & 69 \\
22 & 259.6490 & 23.10 & 0.00260 & 68.95 & 197.00 & 59 & 69 \\
23 & 260.1760 & 23.20 & 0.00263 & 68.95 & 197.00 & 59 & 69 \\
24 & 177.0580 & 10.86 & 0.00153 & 140.00 & 350.00 & 105 & 125 \\
25 & 310.0020 & 7.49 & 0.00194 & 100.00 & 400.00 & 120 & 140 \\
26 & 311.9100 & 7.50 & 0.00195 & 100.00 & 400.00 & 120 & 140 \\
\toprule
\multicolumn{2}{l|}{$t$} & 1 & 2 & 3 & 4 & 5 & 6 \\
\midrule
\multicolumn{2}{l|}{$L$} & 1700 & 1730 & 1690 & 1700 & 1750 & 1850 \\
\multicolumn{2}{l|}{$S$} & 51 & 52 & 51 & 51 & 52 & 56 \\
\midrule
\multicolumn{2}{l|}{$t$} & 7 & 8 & 9 & 10 & 11 & 12  \\
\midrule
\multicolumn{2}{l|}{$L$} & 2000  & 2430 & 2540 & 2600 & 2670 & 2590  \\
\multicolumn{2}{l|}{$S$} & 60 & 73 & 76 & 78 & 80 & 78 \\
\bottomrule
\end{tabular}
}
\label{tab:26unit_a}
\end{table}

\begin{table}[h]
\centering
\caption{Parameters for the 26unit\_time\_b system.}
\resizebox{0.98\linewidth}{!}{%
\begin{tabular}{c|ccccccc}
\toprule
Unit & $A$ & $B$ & $C$ & $p_{\text{min}}$ & $p_{\text{max}}$ & $R^{\text{up}}$ & $R^{\text{dn}}$ \\
\midrule
1 & 24.3891 & 25.55 & 0.02533 & 2.40 & 12.00 & 2 & 3 \\
2 & 24.4110 & 25.68 & 0.02649 & 2.40 & 12.00 & 2 & 3 \\
3 & 24.6382 & 25.80 & 0.02801 & 2.40 & 12.00 & 2 & 3 \\
4 & 24.7605 & 25.93 & 0.02842 & 2.40 & 12.00 & 2 & 3 \\
5 & 24.8882 & 26.06 & 0.02855 & 2.40 & 12.00 & 2 & 3 \\
6 & 117.7550 & 37.55 & 0.01199 & 4.00 & 20.00 & 4 & 5 \\
7 & 118.1080 & 37.66 & 0.01261 & 4.00 & 20.00 & 4 & 5 \\
8 & 118.4580 & 37.78 & 0.01359 & 4.00 & 20.00 & 4 & 5 \\
9 & 118.8210 & 37.89 & 0.01433 & 4.00 & 20.00 & 4 & 5 \\
10 & 81.1364 & 13.33 & 0.00876 & 15.20 & 76.00 & 15 & 19 \\
11 & 81.2980 & 13.36 & 0.00895 & 15.20 & 76.00 & 15 & 19 \\
12 & 81.4641 & 13.38 & 0.00910 & 15.20 & 76.00 & 15 & 19 \\
13 & 81.6259 & 13.41 & 0.00932 & 15.20 & 76.00 & 15 & 19 \\
14 & 217.8950 & 18.00 & 0.00623 & 25.00 & 100.00 & 20 & 25 \\
15 & 218.3350 & 18.10 & 0.00612 & 25.00 & 100.00 & 20 & 25 \\
16 & 218.7750 & 18.20 & 0.00598 & 25.00 & 100.00 & 20 & 25 \\
17 & 142.7350 & 10.69 & 0.00463 & 54.25 & 155.00 & 31 & 39 \\
18 & 142.0290 & 10.72 & 0.00473 & 54.25 & 155.00 & 31 & 39 \\
19 & 143.3180 & 10.74 & 0.00481 & 54.25 & 155.00 & 31 & 39 \\
20 & 143.5970 & 10.76 & 0.00487 & 54.25 & 155.00 & 31 & 39 \\
21 & 259.1310 & 23.00 & 0.00259 & 68.95 & 197.00 & 39 & 49 \\
22 & 259.6490 & 23.10 & 0.00260 & 68.95 & 197.00 & 39 & 49 \\
23 & 260.1760 & 23.20 & 0.00263 & 68.95 & 197.00 & 39 & 49 \\
24 & 177.0580 & 10.86 & 0.00153 & 140.00 & 350.00 & 70 & 88 \\
25 & 310.0020 & 7.49 & 0.00194 & 100.00 & 400.00 & 80 & 100 \\
26 & 311.9100 & 7.50 & 0.00195 & 100.00 & 400.00 & 80 & 100 \\
\toprule
\multicolumn{2}{l|}{$t$} & 1 & 2 & 3 & 4 & 5 & 6  \\
\midrule
\multicolumn{2}{l|}{$L$}& 1700 & 1730 & 1690 & 1700 & 1750 & 1850 \\
\multicolumn{2}{l|}{$S$} & 85 & 86 & 84 & 85 & 88 & 92  \\
\midrule
\multicolumn{2}{l|}{$t$} & 7 & 8 & 9 & 10 & 11 & 12  \\
\midrule
\multicolumn{2}{l|}{$L$} & 2000 & 2430 & 2540 & 2600 & 2670 & 2590  \\
\multicolumn{2}{l|}{$S$} & 100 & 122 & 127 & 130 & 134 & 130  \\
\bottomrule
\end{tabular}
}
\label{tab:26unit_b}
\end{table}

%% file: ilya-biblio.bib
@String{Computing = "Computing" }

@STRING{IEEE = {Proc. {IEEE}}}

@STRING{MA = {Meteorological Applications}}

@String{Springer = "Springer-Verlag" }

@article{shaydulin2019network,
  title={Network community detection on small quantum computers},
  author={Shaydulin, Ruslan and Ushijima-Mwesigwa, Hayato and Safro, Ilya and Mniszewski, Susan and Alexeev, Yuri},
  journal={Advanced Quantum Technologies},
  volume={2},
  number={9},
  pages={1900029},
  year={2019},
  publisher={Wiley Online Library},
  doi={10.1002/qute.201900029},
  url={https://dx.doi.org/10.1002/qute.201900029}
}

@article{schuld2019evaluating,
  title = {Evaluating analytic gradients on quantum hardware},
  author = {Schuld, Maria and Bergholm, Ville and Gogolin, Christian and Izaac, Josh and Killoran, Nathan},
  journal = {Phys. Rev. A},
  volume = {99},
  issue = {3},
  pages = {032331},
  numpages = {7},
  year = {2019},
  month = {Mar},
  publisher = {American Physical Society},
  url = {https://link.aps.org/doi/10.1103/PhysRevA.99.032331}
}

@BOOK{CPLEX,
  title = {Using the {CPLEX} Callable Library},
  publisher = {CPLEX Optimization, Inc.: Incline Village, NV},
  year = {1994},
  author = {{CPLEX Optimization, Inc.}}
}

@INCOLLECTION{N,
  author = {Dontchev, A. L.},
  title = {Lipschitzian stability of {Newton}'s method for variational inclusions},
  booktitle = {System modeling and optimization (Cambridge, 1999)},
  publisher = {Kluwer Acad. Publ.},
  year = {2000},
  pages = {119--147},
  address = {Boston, MA}
}

@INPROCEEDINGS{ramp,
  author = {Tan, Zhangxi and Waterman, Andrew and Avizienis, Rimas and Lee, Yunsup
	and Cook, Henry and Patterson, David and Asanovi\'{c}, Krste},
  title = {RAMP gold: an FPGA-based architecture simulator for multiprocessors},
  booktitle = {DAC},
  year = {2010}
}

@InProceedings{angone2023hybrid,
  author       = {Angone, Anthony and Liu, Xiaoyuan and Shaydulin, Ruslan and Safro, Ilya},
  booktitle    = {2023 IEEE High Performance Extreme Computing Conference (HPEC)},
  title        = {Hybrid quantum-classical multilevel approach for maximum cuts on graphs},
  organization = {IEEE},
  pages        = {1--7},
  year         = {2023},
}

@Article{galda2023similarity,
  author   = {Galda, Alexey and Gupta, Eesh and Falla, Jose and Liu, Xiaoyuan and Lykov, Danylo and Alexeev, Yuri and Safro, Ilya},
  journal  = {Frontiers in Quantum Science and Technology},
  title    = {Similarity-based parameter transferability in the quantum approximate optimization algorithm},
  year     = {2023},
  issn     = {2813-2181},
  volume   = {2},
  doi      = {10.3389/frqst.2023.1200975},
  url      = {https://www.frontiersin.org/articles/10.3389/frqst.2023.1200975},
}

@article{herman2023quantum,
  title={Quantum computing for finance},
  author={Herman, Dylan and Googin, Cody and Liu, Xiaoyuan and Sun, Yue and Galda, Alexey and Safro, Ilya and Pistoia, Marco and Alexeev, Yuri},
  journal={Nature Reviews Physics},
  volume={5},
  number={8},
  pages={450--465},
  year={2023},
  publisher={Nature Publishing Group UK London}
}

@article{salgado2024hybrid,
  title={A hybrid classical-quantum approach to highly constrained Unit Commitment problems},
  author={Salgado, Bruna and Sequeira, Andr{\'e} and Santos, Luis Paulo},
  journal={arXiv preprint arXiv:2412.11312},
  year={2024}
}

@INPROCEEDINGS{bach2024mlqaoa,
  author={Bach, Bao and Falla, Jose and Safro, Ilya},
  booktitle={2024 IEEE International Conference on Quantum Computing and Engineering (QCE)}, 
  title={MLQAOA: Graph Learning Accelerated Hybrid Quantum-Classical Multilevel QAOA}, 
  year={2024},
  volume={01},
  number={},
  pages={1-12},
  doi={10.1109/QCE60285.2024.00072}
}

@InProceedings{maciejewski2024multilevel,
  author       = {Maciejewski, Filip B and Bach, Bao G and Dupont, Maxime and Lott, P Aaron and Sundar, Bhuvanesh and Neira, David E Bernal and Safro, Ilya and Venturelli, Davide},
  booktitle    = {2024 IEEE High Performance Extreme Computing Conference (HPEC)},
  title        = {A multilevel approach for solving large-scale qubo problems with noisy hybrid quantum approximate optimization},
  organization = {IEEE},
  pages        = {1--10},
  year         = {2024},
}

@Article{falla2024graph,
  author    = {Falla, Jose and Langfitt, Quinn and Alexeev, Yuri and Safro, Ilya},
  title     = {Graph representation learning for parameter transferability in quantum approximate optimization algorithm},
  number    = {2},
  pages     = {46},
  volume    = {6},
  journal   = {Quantum Machine Intelligence},
  publisher = {Springer},
  year      = {2024},
}

@Article{tyagin2025qaoa,
  author  = {Tyagin, Ilya and Farag, Marwa H and Sherbert, Kyle and Shirali, Karunya and Alexeev, Yuri and Safro, Ilya},
  title   = {{QAOA-GPT: Efficient Generation of Adaptive and Regular Quantum Approximate Optimization Algorithm Circuits}},
  journal = {accepted in IEEE Quantum Computing and Engineering, arXiv preprint arXiv:2504.16350},
  year    = {2025},
}

@Article{kumar2024recent,
  author    = {Kumar, Gautam and Yadav, Sahil and Mukherjee, Aniruddha and Hassija, Vikas and Guizani, Mohsen},
  title     = {Recent advances in quantum computing for drug discovery and development},
  pages     = {64491--64509},
  volume    = {12},
  journal   = {IEEE Access},
  publisher = {IEEE},
  year      = {2024},
}


%% file: main.bib
@misc{koretsky2021,
      title={Adapting Quantum Approximation Optimization Algorithm (QAOA) for Unit Commitment}, 
      author={Samantha Koretsky and Pranav Gokhale and Jonathan M. Baker and Joshua Viszlai and Honghao Zheng and Niroj Gurung and Ryan Burg and Esa Aleksi Paaso and Amin Khodaei and Rozhin Eskandarpour and Frederic T. Chong},
      year={2021},
      eprint={2110.12624},
      archivePrefix={arXiv},
      primaryClass={quant-ph},
      url={https://arxiv.org/abs/2110.12624}, 
}

@misc{aboumrad2025,
      title={A New Hybrid Quantum-Classical Algorithm for Solving the Unit Commitment Problem}, 
      author={Willie Aboumrad and Phani R V Marthi and Suman Debnath and Martin Roetteler and Evgeny Epifanovsky},
      year={2025},
      eprint={2505.00145},
      archivePrefix={arXiv},
      primaryClass={quant-ph},
      url={https://arxiv.org/abs/2505.00145}, 
}

@article{Sciorilli_2025,
   title={Towards large-scale quantum optimization solvers with few qubits},
   volume={16},
   ISSN={2041-1723},
   url={http://dx.doi.org/10.1038/s41467-024-55346-z},
   DOI={10.1038/s41467-024-55346-z},
   number={1},
   journal={Nature Communications},
   publisher={Springer Science and Business Media LLC},
   author={Sciorilli, Marco and Borges, Lucas and Patti, Taylor L. and García-Martín, Diego and Camilo, Giancarlo and Anandkumar, Anima and Aolita, Leandro},
   year={2025},
   month=jan }

@article{dokucomparative,
  title={A Comparative Study of Hybrid Quantum--Classical Algorithms for the Deterministic Unit Commitment Problem},
  author={Doku, Selase E},
  year={2025}
}

@inproceedings{guan2003optimization,
  title={Optimization based methods for unit commitment: Lagrangian relaxation versus general mixed integer programming},
  author={Guan, Xiaohong and Zhai, Qiaozhu and Papalexopoulos, Alex},
  booktitle={2003 IEEE power engineering society general meeting (IEEE cat. no. 03CH37491)},
  volume={2},
  pages={1095--1100},
  year={2003},
  organization={IEEE}
}

@article{bendotti2019complexity,
  title={On the complexity of the unit commitment problem},
  author={Bendotti, Pascale and Fouilhoux, Pierre and Rottner, C{\'e}cile},
  journal={Annals of Operations Research},
  volume={274},
  number={1},
  pages={119--130},
  year={2019},
  publisher={Springer}
}

@article{osqp,
  author  = {Stellato, B. and Banjac, G. and Goulart, P. and Bemporad, A. and Boyd, S.},
  title   = {{OSQP}: an operator splitting solver for quadratic programs},
  journal = {Mathematical Programming Computation},
  year    = {2020},
  volume  = {12},
  number  = {4},
  pages   = {637--672},
  doi     = {10.1007/s12532-020-00179-2},
  url     = {https://doi.org/10.1007/s12532-020-00179-2},
}

@article{knueven2020mixed,
  title={On mixed-integer programming formulations for the unit commitment problem},
  author={Knueven, Bernard and Ostrowski, James and Watson, Jean-Paul},
  journal={INFORMS Journal on Computing},
  volume={32},
  number={4},
  pages={857--876},
  year={2020},
  publisher={INFORMS}
}

@article{carrion2006computationally,
  title={A computationally efficient mixed-integer linear formulation for the thermal unit commitment problem},
  author={Carri{\'o}n, Miguel and Arroyo, Jos{\'e} M},
  journal={IEEE Transactions on power systems},
  volume={21},
  number={3},
  pages={1371--1378},
  year={2006},
  publisher={IEEE}
}

@inproceedings{chang2004practical,
  title={A practical mixed integer linear programming based approach for unit commitment},
  author={Chang, Gary W and Tsai, YD and Lai, CY and Chung, JS},
  booktitle={IEEE Power Engineering Society General Meeting, 2004.},
  pages={221--225},
  year={2004},
  organization={IEEE}
}

@article{li2005price,
  title={Price-based unit commitment: A case of Lagrangian relaxation versus mixed integer programming},
  author={Li, Tao and Shahidehpour, Mohammad},
  journal={IEEE transactions on power systems},
  volume={20},
  number={4},
  pages={2015--2025},
  year={2005},
  publisher={IEEE}
}

@article{ostrowski2011tight,
  title={Tight mixed integer linear programming formulations for the unit commitment problem},
  author={Ostrowski, James and Anjos, Miguel F and Vannelli, Anthony},
  journal={IEEE transactions on power systems},
  volume={27},
  number={1},
  pages={39--46},
  year={2011},
  publisher={IEEE}
}

@article{putz2021comparison,
  title={A comparison between mixed-integer linear programming and dynamic programming with state prediction as novelty for solving unit commitment},
  author={Putz, Dominik and Schwabeneder, Daniel and Auer, Hans and Fina, Bernadette},
  journal={International Journal of Electrical Power \& Energy Systems},
  volume={125},
  pages={106426},
  year={2021},
  publisher={Elsevier}
}

@article{fan2002new,
  title={A new method for unit commitment with ramping constraints},
  author={Fan, Wei and Guan, Xiaohong and Zhai, Qiaozhu},
  journal={Electric Power Systems Research},
  volume={62},
  number={3},
  pages={215--224},
  year={2002},
  publisher={Elsevier}
}

@article{Mitarai_2018,
   title={Quantum circuit learning},
   volume={98},
   ISSN={2469-9934},
   url={http://dx.doi.org/10.1103/PhysRevA.98.032309},
   DOI={10.1103/physreva.98.032309},
   number={3},
   journal={Physical Review A},
   publisher={American Physical Society (APS)},
   author={Mitarai, K. and Negoro, M. and Kitagawa, M. and Fujii, K.},
   year={2018},
   month=sep }

@article{Schuld_2019,
   title={Evaluating analytic gradients on quantum hardware},
   volume={99},
   ISSN={2469-9934},
   url={http://dx.doi.org/10.1103/PhysRevA.99.032331},
   DOI={10.1103/physreva.99.032331},
   number={3},
   journal={Physical Review A},
   publisher={American Physical Society (APS)},
   author={Schuld, Maria and Bergholm, Ville and Gogolin, Christian and Izaac, Josh and Killoran, Nathan},
   year={2019},
   month=mar }

@article{cplex2022v22,
  title={IBM ILOG CPLEX Optimization Studio v22.1},
  author={IBM},
  journal={International Business Machines Corporation},
  year={2022}
}

@misc{qiskitqce,
  title={Pauli Correlation Encoding to reduce Maxcut requirements},
  url={https://quantum.cloud.ibm.com/docs/en/tutorials/pauli-correlation-encoding-for-qaoa},
  journal={International Business Machines Corporation},
  urldate={2026-03-13},
}

@article{colson2007overview,
  title   = {An Overview of Bilevel Optimization},
  author  = {Colson, Beno{\^\i}t and Marcotte, Patrice and Savard, Gilles},
  journal = {Annals of Operations Research},
  year    = {2007},
  volume  = {153},
  number  = {1},
  pages   = {235--256},
  doi     = {10.1007/s10479-007-0176-2}
}

@book{dempe2002foundations,
  title     = {Foundations of Bilevel Programming},
  author    = {Dempe, Stephan},
  publisher = {Springer},
  year      = {2002},
  doi       = {10.1007/b101970}
}

@book{bard1998practical,
  title     = {Practical Bilevel Optimization: Algorithms and Applications},
  author    = {Bard, Jonathan F.},
  publisher = {Springer},
  year      = {1998},
  doi       = {10.1007/978-1-4757-2836-1}
}

@inproceedings{amos2017optnet,
  title={Optnet: Differentiable optimization as a layer in neural networks},
  author={Amos, Brandon and Kolter, J Zico},
  booktitle={International conference on machine learning},
  pages={136--145},
  year={2017},
  organization={PMLR}
}

@article{agrawal2019differentiable,
  title={Differentiable convex optimization layers},
  author={Agrawal, Akshay and Amos, Brandon and Barratt, Shane and Boyd, Stephen and Diamond, Steven and Kolter, J Zico},
  journal={Advances in neural information processing systems},
  volume={32},
  year={2019}
}

@book{bonnans2013perturbation,
  title={Perturbation analysis of optimization problems},
  author={Bonnans, J Fr{\'e}d{\'e}ric and Shapiro, Alexander},
  year={2013},
  publisher={Springer Science \& Business Media}
}

@article{crooks2019gradients,
  title={Gradients of parameterized quantum gates using the parameter-shift rule and gate decomposition},
  author={Crooks, Gavin E},
  journal={arXiv preprint arXiv:1905.13311},
  year={2019}
}

@article{cherrat2024quantum,
  title={Quantum vision transformers},
  author={Cherrat, El Amine and Kerenidis, Iordanis and Mathur, Natansh and Landman, Jonas and Strahm, Martin and Li, Yun Yvonna},
  journal={Quantum},
  volume={8},
  number={arXiv: 2209.08167},
  pages={1265},
  year={2024}
}

@misc{efficientsu2,
  title={EfficientSU2},
  url={https://quantum.cloud.ibm.com/docs/en/api/qiskit/qiskit.circuit.library.EfficientSU2},
  journal={International Business Machines Corporation},
  urldate={2026-04-27},
}

@article{soloviev2025large,
  title={Large-scale portfolio optimization using Pauli Correlation Encoding},
  author={Soloviev, Vicente P and Krompiec, Michal},
  journal={arXiv preprint arXiv:2511.21305},
  year={2025}
}

@article{padin2026pauli,
  title={Pauli Correlation Encoding for Budget-Constrained Optimization},
  author={Pad{\'\i}n-Mart{\'\i}nez, Jacobo and Soloviev, Vicente P and Borrallo-Rentero, Alejandro and Rodr{\'\i}guez-Otero, Ant{\'o}n and Alfonso-Rodr{\'\i}guez, Raquel and Krompiec, Michal},
  journal={arXiv preprint arXiv:2602.17479},
  year={2026}
}

@article{do2026warm,
  title={Warm-Starting PCE for Traveling Salesman Problem},
  author={do Carmo, Rafael Sim{\~o}es and dos Reis, Renato Gomes and F Silva, Samuel Fernando and E Arruda, Luiz Gustavo and Fanchini, Felipe F and others},
  journal={Brazilian Journal of Physics},
  volume={56},
  number={1},
  pages={49},
  year={2026},
  publisher={Springer}
}
